\documentstyle[preprint,prl,aps]{revtex}
\begin{document}
\draft

\title{
Spin Gaps in a Frustrated Heisenberg model for CaV$_4$O$_9$
}
\author{ Steven R.\ White}
\address{ 
Department of Physics and Astronomy,
University of California,
Irvine, CA 92717
}
\date{\today}
\maketitle
\begin{abstract}
I report results of a density matrix renormalization group (DMRG) study
of a model for the two dimensional spin-gapped system CaV$_4$O$_9$.
This study represents the first time that DMRG has been
used to study a two dimensional system on large lattices, in this
case as large as $24\times 11$, allowing extrapolation to the
thermodynamic limit. I present a substantial improvement
to the DMRG algorithms which makes these calculations feasible.
\end{abstract}

\narrowtext

Since the discovery of the high temperature superconductivity,
condensed matter physicists have searched for other two dimensional
systems with exotic ``spin liquid'' ground states. Thus considerable 
excitement accompanied the recent discovery of CaV$_4$O$_9$, a
two dimensional, frustrated $S=1/2$ Heisenberg spin system, with
a substantial spin gap\cite{taniguchi}.
This system has been modeled by a depleted square lattice Heisenberg
model, with both nearest and next-nearest exchange
interactions\cite{ueda}. It consists of a square lattice
with $1/5$ of the spins missing, as shown in Figure 1, and has been
called a CAVO lattice\cite{gelfand}. It is believed that the
superexchange is mediated by out-of-plane oxygen atoms, resulting
in a very large next-nearest exchange: $J'\approx J/2$\cite{ueda}. 
This frustrating interaction helps stabilize
the spin-gapped state against N\'eel order.  

One can think of the ground state of this system
as a ``plaquette resonating valence bond'' state\cite{ueda}:
no phase transition is expected if the interactions between
plaquettes are adiabatically removed, and the ground state of
a single plaquette is perfectly described by a resonating valence
bond (RVB) variational ansatz (in the absence of frustration).
In addition, in the weakly-interacting plaquette limit, pairs of
holes bind on plaquettes, suggesting the possibility of a
superconducting ground state upon doping. The system is
reminiscent of ladder systems with even numbers of legs. In that
case, somewhat more complicated RVB states have been 
useful in describing the spin liquid ground states of undoped 
ladders\cite{rice,rvbprl}, and upon doping, strong pairing
correlations are observed numerically\cite{noack}. 

A number of theoretical and numerical treatments have been performed on this
model in the last year\cite{ueda,katoh,sano,mila,troyer,starykh}.
Troyer, Kontani and Ueda made the most reliable determination 
of the {\it unfrustrated } phase diagram using a quantum Monte Carlo 
loop algorithm\cite{troyer}. 
An important conclusion of this study was that simple $1/5$ depletion of
the isotropic square lattice, without frustration, does {\it not }
destroy N\'eel order. A spin liquid ground state {\it was } found when 
the couplings within a plaquette were about 10\% greater
than between plaquettes. This result contradicted earlier (non-loop)
quantum Monte Carlo calculations on smaller systems\cite{katoh}.
It was not possible to include frustration because of sign problems.
Gelfand, et. al.\cite{gelfand} applied series expansion techniques
and were able to study the frustrated and unfrustrated systems.
Their results were in agreement with Troyer, et. al. for the
unfrustrated system. They concluded that a next-nearest neighbor
interaction $J'=J/2$ was consistent with experimental results.

I present results here from density matrix renormalization group (DMRG) 
calculations\cite{DMRG} for the spin gap of the frustrated CAVO lattice.
The results are in agreement with those of Gelfand, et. al., and, 
in fact, when extrapolated to the thermodynamic limit, appear 
to be more accurate. Although DMRG is usually much
more accurate than other numerical techniques for large one
dimensional systems, these are the first reliable results for 
systems wide enough to be considered two dimensional---up to
$24\times 11$. These calculations are feasible because of an
important improvement to 
the DMRG algorithms, which I present here, which increases the
speed of the calculations by up to two orders of magnitude.

The improvement to DMRG involves keeping track of the wavefunction from
step to step.  The step referred to here is the process of adding
a site to a block and requires the diagonalization of a superblock
configuration of two blocks and two sites\cite{DMRG}. In each DMRG step, 
an iterative sparse matrix algorithm, such as the Davidson method,
is used to find the ground state of the superblock.
In the original formulation of DMRG, no
starting point for the Davidson procedure was specified. 
To ensure that the DMRG procedure is always stable and convergent, 
the superblock ground state usually has to be determined to rather 
high accuracy.  (One diagonalization which converges to a
low-lying eigenstate other than the ground state ruins the accuracy
of the entire DMRG calculation.) Consequently, a substantial number of 
Davidson steps
are necessary to converge to sufficient accuracy, typically 40-100. 
The total calculation time is proportional to the average number
of Davidson steps.

If a very good initial guess is available for the Davidson procedure,
the number of Davidson steps can be reduced substantially.
An ideal initial guess, for the case of the finite system DMRG
algorithm, is the final wavefunction from the previous DMRG step.  
This wavefunction, however, is in a different basis, 
corresponding to a different superblock, but it can be transformed into
the basis corresponding to the current superblock, as I describe
below. 
Use of this transformation to obtain the initial state in a Davidson
diagonalization can reduce the number of Davidson steps by one half,
typically, assuming that one iterates Davidson until it converges
to high accuracy. 
Use of this initial guess has an even more important
advantage: it is not necessary to converge to high accuracy, since
there is no danger of converging to an incorrect low-lying 
eigenstate. The initial guess not only has low energy, it
approximately describes
the correct eigenstate, as obtained in the previous step. In fact,
the algorithm can be made completely stable even if the number of
Davidson steps is restricted to two or three! Thus one saves a
factor of 20-50 in the time required by the Davidson procedure.
The overall speedup is somewhat reduced from this factor because 
the calculation
time to perform other parts of the DMRG procedure, such as
diagonalizing the density matrix, becomes significant.

A DMRG step adds a site onto a block, constructing an appropriate
basis for the new block. Let $|\alpha_{l}\rangle$ be the states of
left block $l$, where $l$ is the rightmost site of the block.
 Let $|s_{l}\rangle$ be the states of site $l$.  Then the basis states 
for the new left block are given by
\begin{equation}
|\alpha_{l+1}\rangle = \sum_{s_{l+1},\alpha_{l}} 
L^{l+1}[s_{l+1}]_{\alpha_{l+1},\alpha_{l}} 
|\alpha_{l}\rangle \otimes |s_{l+1}\rangle .
\end{equation}
This notation is similar to that of Ostlund and Rommer\cite{ostlund}.
The transformation matrix
$L^{l+1}[s_{l+1}]_{\alpha_{l+1},\alpha_{l}}$ 
is a slightly rewritten form of the truncated matrix of density matrix 
eigenvectors. 
The states of the right block $|\beta_{l+3}\rangle$ were formed at an 
earlier DMRG step in a similar fashion
\begin{equation}
|\beta_{l+3}\rangle = \sum_{s_{l+3},\beta_{l+4}} 
R^{l+3}[s_{l+3}]_{\beta_{l+3},\beta_{l+4}} |s_{l+3}\rangle \otimes 
|\beta_{l+4}\rangle .
\end{equation}
I do not assume any reflection symmetry for the lattice: the $L$
and $R$ matrices are independent.

A superblock basis state is written in the form 
\begin{equation}
|\alpha_{l}s_{l+1}s_{l+2}\beta_{l+3}\rangle =
|\alpha_{l}\rangle \otimes |s_{l+1}\rangle \otimes 
|s_{l+2}\rangle \otimes |\beta_{l+3}\rangle .
\end{equation}
A superblock wavefunction $|\psi\rangle$ is written in this 
basis as
\begin{equation}
|\psi\rangle = \sum_{\alpha_{l}s_{l+1}s_{l+2}\beta_{l+3}}
\psi(\alpha_{l}s_{l+1}s_{l+2}\beta_{l+3})
|\alpha_{l}s_{l+1}s_{l+2}\beta_{l+3}\rangle .
\end{equation}
One needs to transform this wavefunction into the basis appropriate
for the next DMRG step,
$|\alpha_{l+1}s_{l+2}s_{l+3}\beta_{l+4}\rangle$.
The transformation is not exact, since there is a truncation in
going from $|\alpha_{l}s_{l+1}\rangle$ to $|\alpha_{l+1}\rangle$.
However, the states $|\alpha_{l+1}\rangle$ are formed using the
density matrix to be ideally adapted for representing
$|\psi\rangle$, so for the transformation of the wavefunction
only, one can approximate
\begin{equation}
\sum_{\alpha_{l+1}} |\alpha_{l+1} \rangle \langle \alpha_{l+1} | \approx 1 .
\end{equation}

With this approximation one readily obtains
\begin{equation}
\psi(\alpha_{l+1}s_{l+2}s_{l+3}\beta_{l+4}) \approx
\sum_{\alpha_{l}s_{l+1}\beta_{l+3}} 
L^{l+1}[s_{l+1}]_{\alpha_{l+1},\alpha_{l}}
\psi(\alpha_{l}s_{l+1}s_{l+2}\beta_{l+3})
R^{l+3}[s_{l+3}]_{\beta_{l+3},\beta_{l+4}} .
\end{equation}
The most efficient way to implement this transformation numerically is
to first form the intermediate wavefunction
\begin{equation}
\psi(\alpha_{l+1}s_{l+2}\beta_{l+3}) =
\sum_{\alpha_{l}s_{l+1}} 
L^{l+1}[s_{l+1}]_{\alpha_{l+1},\alpha_{l}}
\psi(\alpha_{l}s_{l+1}s_{l+2}\beta_{l+3}) ,
\end{equation}
and then form the final result
\begin{equation}
\psi(\alpha_{l+1}s_{l+2}s_{l+3}\beta_{l+4}) =
\sum_{\beta_{l+3}} 
\psi(\alpha_{l+1}s_{l+2}\beta_{l+3})
R^{l+3}[s_{l+3}]_{\beta_{l+3},\beta_{l+4}} .
\end{equation}
In this form, the transformation requires very computer little time compared to
other parts of the calculation.

This transformation is used for one half of the DMRG steps, when a site
is being added to the left block. An analogous transformation is
used for adding a site to the right block.

Implementing this transformation requires saving all the matrices $L$ and
$R$, which is ordinarily not done. The storage for these matrices is
typically 20-30\% of the storage required for the blocks themselves,
so the extra storage is not a major concern. In an efficient DMRG
implementation for a typical machine, such as a Cray or a workstation, 
both the blocks and the transformation matrices should be stored on
disk. The calculations described here sometimes required more than
a gigabyte of scratch disk storage, but never more than 80-90 megabytes
of RAM.

In many one dimensional systems, DMRG converges in one or two
sweeps through the system. In quasi-one or two dimensional systems,
the number of sweeps needed can easily grow to five to 10.
Another important improvement in efficiency comes from gradually
increasing the number of states kept per block as one performs
the sweeps\cite{noackchap}. In this case the calculation time is 
dominated by the last sweep, during which the number of Davidson 
steps per DMRG step can be constrained to be only two or three. 
Compared to a DMRG calculation keeping a constant number of 
states and without the wavefunction transformation, the speedup 
can be over two orders of magnitude.

If the lattice is reflection-symmetric, an additional factor of
two can be saved in both calculation time and memory, but 
a different wavefunction transformation is needed for the DMRG step where
the superblock is symmetric.
Also, it is possible to adopt somewhat similar methods to obtain good
initial guesses for the wavefunction in the infinite system method.
These techniques will be reported elsewhere\cite{tobepub}.

The beginning of any DMRG calculation of a 2D system is a mapping
of the 2D lattice onto a 1D chain---basically, one must choose an
order to traverse the sites. It is standard to use the scanline
mapping---fix $x$, step through all values of $y$, then increment
$x$, etc.---which has the advantage of keeping the blocks as contiguous
as possible. In treating the CAVO lattice, 
it is more advantageous to modify this
slightly so that all the sites in a plaquette are traversed in succession. 
This incorporates the fact that for the parameters I consider here,
correlations within a plaquette are strongest.

I consider the Heisenberg Hamiltonian
\begin{equation}
H = J_{ij} \sum_{\langle i,j \rangle} {\hbox{\bf S}}_{i} \cdot {\hbox{\bf
S}}_{j}
\end{equation}
defined on an $L_x\times L_y$ CAVO lattice with $S=\frac12$.
As shown in Fig. 1,
I take all nearest-neighbor $J_{ij}$ to be identical, with value
$J_1 = 1$, setting the energy scale. 
All next-nearest-neighbor $J_{ij}$ are also identical, with
value $J_2$. All other $J_{ij}$ are zero.

I have studied open systems, although putting periodic boundary
conditions in the short direction $y$ is not particularly difficult.
Having open boundary conditions allows a variety of sizes to be
studied. We consider a set of systems of length 24 and width up to
11. Since interactions are strongest within plaquettes, only full
plaquettes are included. 
We keep up to $m=600$ states per block, with truncation error at worst
about $10^{-5}$. Typical errors in the total energy, for the larger
systems, were less than about $10^{-3}$. 
For systems of width 8 and larger, an
extrapolation to $m \to \infty$ was used, assuming an exponential
fall off in the error in the energy as a function of
$m$\cite{whitehuse,zigzag}. 
Corrections were at most about $10^{-3}$.
For each system we calculate the ground state
energies with quantum numbers $S_z=0$ and $S_z=1$. The spin gap
$\Delta$
is the difference in energies. The largest system, $24\times11$,
took about 15 - 20 hours of workstation time (rated at 135 SPECfp92)
for one value of $S_z$.

Figure 2 shows some of the results, for various widths of the system
$L_y$. From the width 7 data, we see
that the spin gap is peaked at $J_2=0.5$, which also happens to
be appropriate for CaV$_4$O$_9$.  For larger values
of $J_2$ it falls rapidly towards zero. (The gap for $J_2=0.8$ was
consistent with zero, within uncharacteristically large error bars
of about 0.05.)

Finite size extrapolation is crucial to determine the spin gap
for smaller values of $J_2$. Excellent extrapolations can be
obtained if one assumes the low lying spin excitations obey
a {\it relativistic} dispersion relation
\begin{equation}
\Delta(k)^2 = \Delta^2 + v^2 k^2 ,
\end{equation}
where $\Delta$ is the bulk gap and the velocity $v$ corresponds
to the speed of light.
Lorentz invariant low energy excitations are common (but not
universal) in gapped,
one dimensional spin systems, reflecting Lorentz invariance
of the corresponding nonlinear sigma model. Considerations of
simple particle-in-a-box systems with various boundary conditions
indicate that a generic boundary condition at the edge of an
open system fixes the logarithmic derivative of the wavefunction
$\psi$, $\frac{d\psi}{dx}/\psi = \hbox{const}$ . This can be shown to
imply that the lowest value of $k$ allowed in a 1-D box of size $L$
is given by $\pi/(L-a)$, where $a$ depends on boundary effects but
is independent of $L$. This
leads to the following form for the gap as a function of system size
\begin{equation}
\Delta(L_x,L_y)^2 = \Delta^2 + \frac{\pi^2v^2}{(L_x-a)^2} +
\frac{\pi^2v^2}{(L_y-a)^2}.
\end{equation}
I have found that for $J_2< \ \sim 0.3$, we can set $a=0$, and
still obtain excellent fits. This is indicated by linear behavior
when $\Delta^2$ is plotted versus $1/L_y^2$, with $L_x$ fixed.
Results are shown in Figure 3 for typical values of $J_2$.
For $J_2=0.5$, the fits were poor for $a=0$, and
mediocre with $a$ nonzero, because the data was slightly irregular.
Gelfand, et.al.\cite{gelfand} observed that the
gap minimum can move away from
$(\pi,\pi)$ for larger values of $J_2$. This data supports that
proposition for $J_2=0.5$. In the case of an incommensurate gap
minimum, we would expect irregular behavior of the gap as a
function of $L$, as the value of $k$ allowed by the lattice 
which is closest to the minimum would jump about as $L$ increased.

Using the fits shown, I corrected for the finite value of $L_x$ and
obtained results for $\Delta$ in the thermodynamic limit. 
Figure 4 shows the results as a function of $J_2$.
The extrapolation using Eq. (11) yielded imaginary gaps for 
$J_2 = 0$ and $0.05$, which we interpret to mean $\Delta=0$.
The transition to a spin gapped state appears
at $J_2 = 0.06(1)$. This result is in agreement with previous
work indicating that the system with $J_2=0$ is close to the
disordered phase. The value at $J_2=0.5$, $\Delta = 0.515(15)$,
is somewhat lower than the series results of Gelfand, et.al.,
$\Delta = 0.57(3)$. However, the results are completely consistent
if the shift in the gap minimum results in an overestimate in
the series results by about 0.05, as Gelfand, et.al. suggest.

I have presented the first results using DMRG on a two dimensional 
system for lattices wide enough to allow extrapolation to the 
thermodynamic limit.
DMRG still is primarily a one dimensional technique, in that
the accuracy falls off rapidly as the system's width increases.
The CAVO system studied here was less difficult than many other
two dimensional systems, both because of the existence
of a gap and the depleted character of the lattice. Nevertheless, 
as DMRG and computational resources improve, I expect it will 
become a standard numerical technique for two dimensional systems,
including doped fermion systems.

I thank Naokazu Shibata and Rajiv Singh for helpful conversations.
I acknowledge support from the Office of Naval Research under
grant No. N00014-91-J-1143, and
from the NSF under Grant No.\ DMR-9509945.
Some of the calculations were performed at the
San Diego Supercomputer Center.

\begin{figure}
\caption{A $12\times7$ open CAVO lattice. The solid lines are
nearest-neighbor bonds with exchange $J_1$, and the dotted lines
are next-nearest-neighbor bonds with exchange $J_2$. Only complete
plaquettes which fit within a $12\times7$ rectangle are retained.
}
\label{a}
\end{figure}
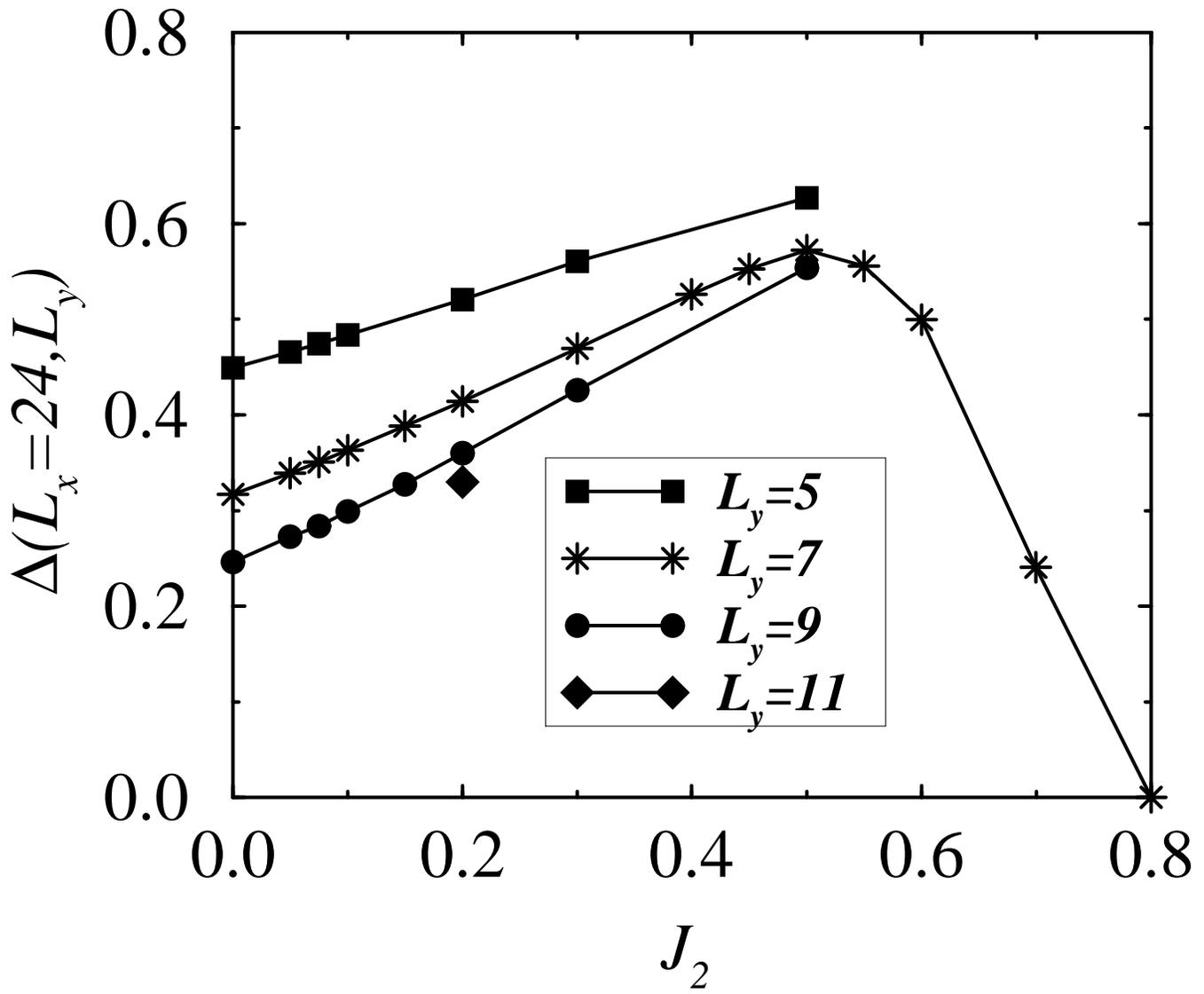
\begin{figure}
\caption{Spin gap as a function of $J_2$ for various widths $L_y$,
with $L_x=24$.
}
\label{b}
\end{figure}
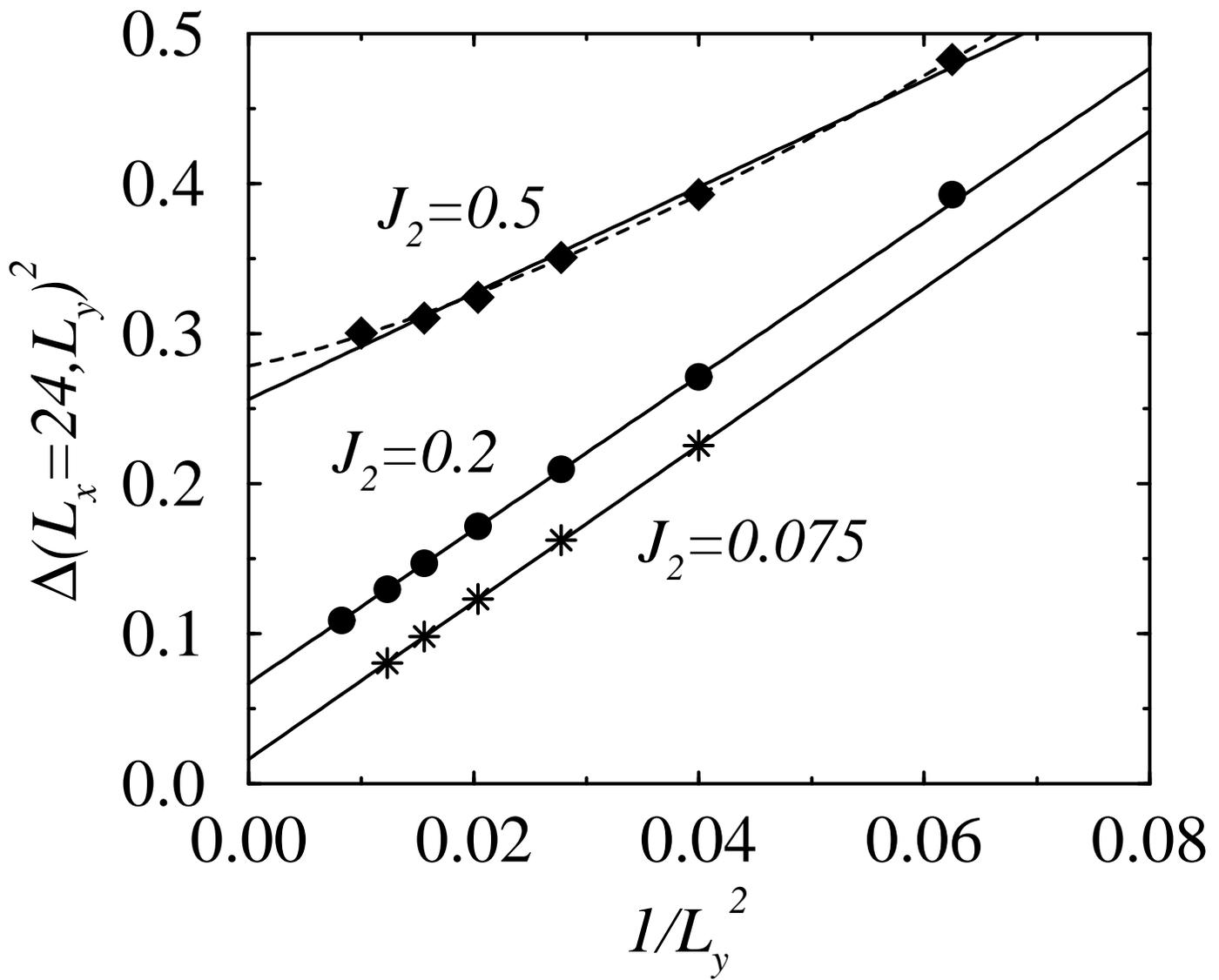
\begin{figure}
\caption{Gap squared as a function of $L_y^{-2}$.
The solid lines are linear fits, excluding
the $L_y=4$ point for $J'=0.2$. The dashed line is an alternative
fit which includes an $L_y^{-3}$ term.
}
\label{c}
\end{figure}
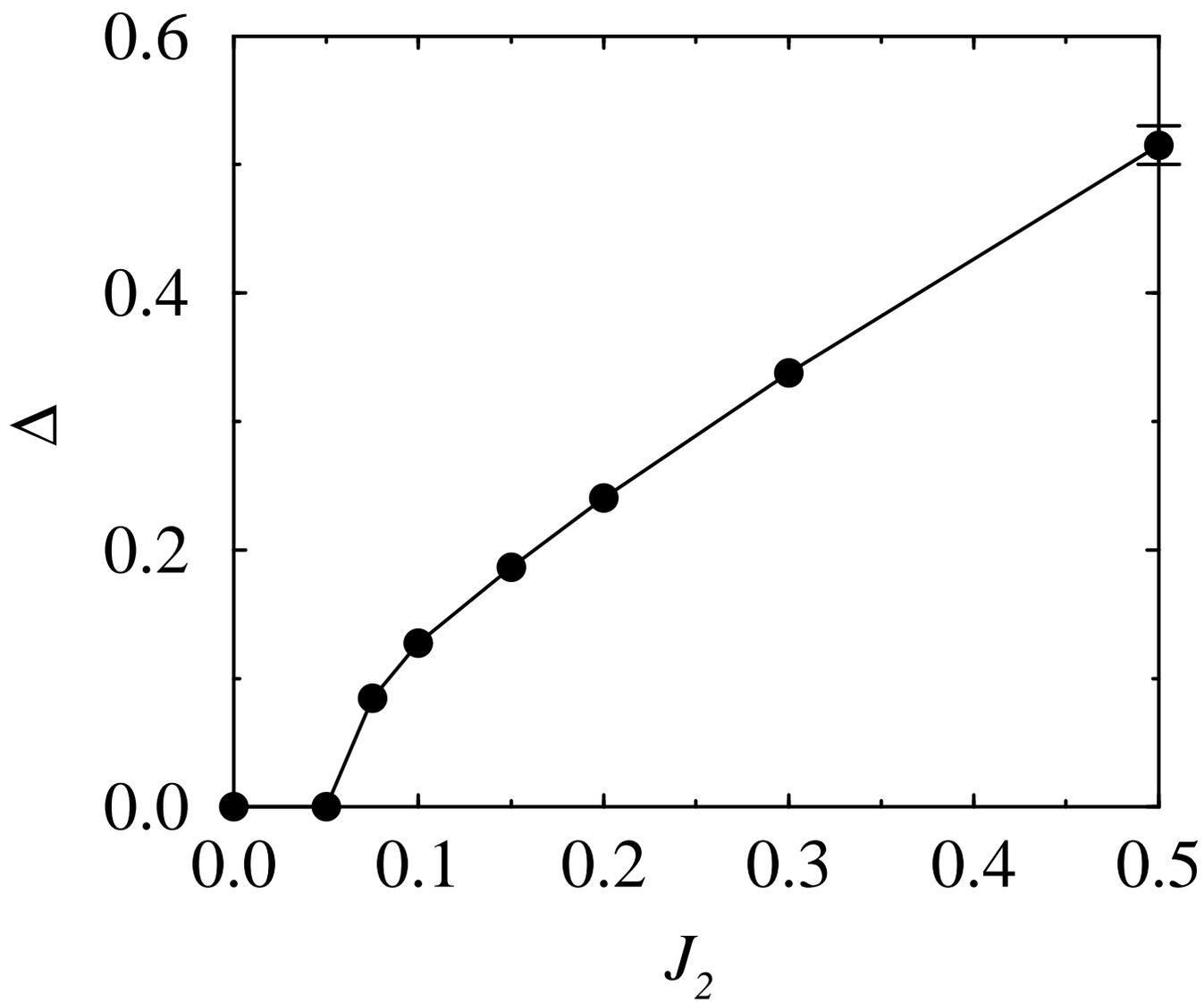
\begin{figure}
\caption{
Spin gap extrapolated to the thermodynamic limit, as a function 
fo $J_2$. Where not shown, errors are comparable to the point
size.
}
\label{d}
\end{figure}

\end{document}